\documentclass[twocolumn]{aastex6}
\usepackage{natbib}
\usepackage{amsmath}
\usepackage{graphicx}
\renewcommand{\phi}{\varphi}
\newcommand{\rhessi}{\emph{RHESSI}\ }

\shorttitle{Escape and Scattering Times of Energetic Particles}
\shortauthors{F.\ Effenberger \& V.\ Petrosian}

\def\tsc{\tau_{\rm sc}\,}

\def\tcross{\tau_{\rm cross}\,}
\def\tesc{T_{\rm esc}\,}

\begin{document}

\received{?}
\accepted{?}

\title{The Relation Between Escape and Scattering Times of Energetic Particles in a Turbulent
Magnetized Plasma: Application to Solar Flares}

\author{Frederic~Effenberger}
\affil{Helmholtz Centre Potsdam, GFZ, German Research Centre for Geosciences, Potsdam, Germany\\ Bay
Area Environmental Research Institute, Petaluma, CA, USA}
\and 
\author{Vah\'e~Petrosian}
\affil{Department of Physics and KIPAC,
  Stanford University, Stanford, CA, USA}

\email{feffen@gfz-potsdam.de, vahep@stanford.edu}

\begin{abstract}
A knowledge of the particle escape time from the acceleration regions of many space and
astrophysical sources is of critical importance in the analysis of emission signatures produced by these particles and in the
determination of the acceleration and transport mechanisms at work. This paper addresses this
general problem, in particular in solar flares, where in addition to scattering by turbulence, the
magnetic field convergence from the acceleration region towards its boundaries 
also influences the particle escape. We test an (approximate) analytic relation between
escape and scattering times, and the field convergence rate, based on the work of Malyshkin and
Kulsrud (2001), valid for both strong and weak diffusion limits and isotropic pitch angle
distribution of the injected particles, with a numerical model of particle transport.
To this end, a kinetic Fokker-Planck transport model of particles is solved with a stochastic differential equation scheme
assuming different initial pitch angle distributions. This approach
enables further insights into the phase-space dynamics of the transport process, which would
otherwise not be accessible. We find that in general the numerical results agree well with the
analytic equation for the isotropic case, however, there are  significant differences 
weak diffusion regime for non-isotopic cases, especially for distributions beamed along the
magnetic field lines. The results are important in the interpretation of observations of energetic
particles in solar flares, and other similar space and astrophysical acceleration sites and for the
determination of acceleration-transport coefficients, commonly used in Fokker-Planck type
kinetic equations.
\end{abstract}


\section{Introduction}
\label{intro}

The processes involved in the acceleration and transport of energetic particles in many space and astrophysical
settings are still a very active topic of investigation after decades of research. These
processes can be investigated by the observations of nonthermal radiations emitted from these sites
and from the spectrum of cosmic rays (CRs) escaping them. Examples of these are solar eruptive events
involving nonthermal radiation produced by flare accelerated particles and solar energetic particles
(SEPs) seen by near Earth instruments. The aim of this paper is to clarify the transport
coefficients involved in the acceleration-transport processes with particular emphasis on the time
the particles spend in the acceleration site, which we refer to as the {\it escape time}, $\tesc$.
We will use solar observations as an example for our discussion.

The escape time is an important component of acceleration-transport process for several
reasons. Clearly, the time spent in the acceleration site is important in shaping the energy, $E$,
spectrum of the particles in the acceleration site, $N(E)$. It is also the main factor determining
the spectrum of the flux of the escaping particles, ${\dot Q}(E)=N(E)/\tesc(E)$. In most sources, the
main transport characteristics that determines the escape time are the crossing, $\tcross=L/v$, and
scattering, $\tsc(E)$, times, for a source of size $L$  and a particle with speed $v$. Scattering
can be due to Coulomb interactions in a collisional plasma and/or wave-particle interactions in a
turbulent plasma. The later is related to the stochastic acceleration rate by turbulence or
the acceleration rate in a shock environment \citep[see, e.g.,][]{Petrosian-2012}. In addition, in
situations with weaker diffusion rate (i.e.~when  $\tsc(E)>\tcross$ due to low particle and
turbulence densities), the
background guiding magnetic field, $B$, can affect the
escape time due to mirroring in a converging field geometry. Thus, for a comprehensive
analysis of particle confinement and 
escape from the acceleration regions, the field convergence
towards the boundaries of the acceleration region
needs to be considered in a
particle transport model. The escape time
(through its relation with $\tsc$) is related to all transport coefficients so that
clarifications of its role can shed
light on many aspects of the acceleration process.
 
In general, in the acceleration process of background 
thermal particles (with a Maxwellian distribution), the interplay between Coulomb
and turbulent scattering usually leads to plasma heating and acceleration, and can also lead
to stable particle distributions consisting of a quasi-thermal components with non-thermal tails,
often described by kappa distributions \citep{Bian-etal-2014}, for which evidence exists from
solar flare observations \citep{Kasparova-Karlicky-2009,Oka-etal-2013,Oka-etal-2015,Oka-etal-2018}.
As shown in \citet{Petrosian-East-2008} and 
\citet{Petrosian-Kang-2015}, in a {\it closed
system}, i.e.\ where the particle escape time is longer than 
all the other timescales, irrespective of the details of
the acceleration process, most of the energy  goes into
heating the plasma rather than producing a nonthermal tail.
But, when the escape time is shorter, then a substantial population of nonthermal particles can escape the
acceleration site; with an spectrum not necessarily the same as that of the
accelerated one. They are distinguished by the escape time. This distinction is important in 
many space and astrophysical accelerators, in particular in solar eruptive events, as
described below. 

A consequence of the particle interactions in the solar atmosphere is the production of thermal
(due to plasma heating) and non-thermal (due to acceleration) radiation, in particular hard X-ray
(HXR) bremsstrahlung, as observed for example with the Reuven Ramaty High Energy Solar Spectroscopic
Imager \citep[\emph{RHESSI},][]{Lin-etal-2002}. The HXR observations by \rhessi (and earlier by {\it
Yohkoh}) have shown the presence of a distinct source near the flaring loop top region (presumably
the acceleration site) produced by the accelerated electrons. This is in addition
to the more prominent foot point emission produced by escaping electrons.
\citep{Masuda-etal-1994,Liu-etal-2008,Krucker-etal-2010,Liu-etal-2013},
which appear to be a common feature of almost all {\it Yohkoh} \citep{Petrosian-etal-2002} 
and {\it RHESSI} \citep{Liu-etal-2006, Krucker-Lin-2008} flares. These two emissions are
related through the escape time.
\citet{Petrosian-Donaghy-1999} showed that this requires
some confinement of the electrons near the loop top acceleration site, which makes the escape
time longer than the crossing time, and proposed
turbulence as the agent of scattering  and acceleration \citep[see also][]{Kontar-etal-2014}.
Coulomb scattering can also trap particles at the loop top if the densities are high.
However, because Coulomb
energy loss and scattering rates are comparable, in such a case electrons lose most of their
energy at the loop top leading to weaker footpoint emission. As shown by \citet{Leach-Petrosian-1983},
with Coulomb collision alone one obtains a gradual decline of emission along the flaring loop with a
rapid increase below the transition region. \citet{Leach-Petrosian-1983} also showed that convergence
of magnetic field toward the photosphere can enhance the trapping of the particles (see their
Fig.~13). These effects were also discussed in \citet{Fletcher-1995} and
\citet{Fletcher-Martens-1998} with similar results. They find that the confinement by the loop
magnetic field can lead to a loop top emission that is stronger than or comparable to the footpoint
emission for densities of $3\times 10^{10} (4\times 10^{9}$) cm$^{-3}$; see
\citet{Fletcher-Martens-1998} Figs.~7 and 9, respectively.

As evident from the above discussion, observations of loop top and footpoint emissions can
provide information on the escape time.
The relation of HXR emission and energetic electron properties can be analyzed with forward
fitting methods or by regularized inversion using the imaging spectroscopy abilities
of \rhessi \citep{Piana-etal-2007}. As shown by
\citet{Petrosian-Chen-2010}, the inversion method allows the determination of the
escape time from the comparison of loop top and footpoint non-thermal
electron images obtained nonparametrically from \rhessi data directly.
Subsequently, \citet{Chen-Petrosian-2013} showed that with this technique in addition to $\tesc$,
one can obtain the other relevant coefficients (energy loss, acceleration and
crossing times). This analysis has provided a paradigm shift indicating that the mirroring
effect can be the main source of confinement of particles in the acceleration site. Further
evidence supporting this results comes from the interpretation by \citet{Petrosian-2016} of
\citet{Krucker-etal-2007} data comparing the spectra of HXR producing and SEP electrons in impulsive,
prompt events. These findings can also be useful in the interpretation of the coronal
emission close to the acceleration sites in partially occulted flares
\citep[e.g.,][]{Krucker-Lin-2008,Effenberger-etal-2016,Effenberger-etal-2017}, with more
direct information on the acceleration process.

Magnetic field convergence and the mirroring effect can also be important in the transport of
particles from coronal mass ejection (CME) shock environments. It is generally accepted that
SEPs observed near the Earth are particles escaping from flare sites or the upstream region of such shocks. Recently,
the {\it Fermi} large area telescope has detected  $>100$ MeV sustained solar gamma ray emission from many eruptive
events \citep{Ajello-etal-2014,Pesce-Rollins-etal-2015,Ackermann-etal-2017} associated with fast CMEs, lasting almost as long as the accompanying SEPs. These
post-impulsive emission, with no other accompanying radiative signatures, have raised the
possibility that they may be produced by particles escaping the turbulent downstream region of the
CME-shock back to the Sun along converging field lines \citep{Jin-etal-2018}. Thus again, analysis of these events
requires a knowledge of the escape time from a region where turbulence and field geometry can play
an important role.

An analytic approximation relating  the escape
and scattering times of particles in a converging field environment has
been provided in \citep{Malyshkin-Kulsrud-2001}. One of our goals is to test the validity of this
relation with a numerical particle transport model and explore different initial pitch-angle distributions, in
addition to the isotropic one considered by these authors.

In the next section we describe the origin of this analytic expression, and in \S 3 we
present the transport equations of particles in
a turbulent site with simple converging field geometry, the simulation scheme we use to
determine pitch angle and spatial distribution of particles subject to only pitch angle scattering,
and address the determination of the escape time. The results are presented and discussed in \S 4 followed by
the summary and conclusions in \S 5.


\section{The Escape Time}
\label{escape}

As described above, several factors play an important role in trapping the
particles and determining how fast particles can escape a turbulent magnetized plasma, which is the
case for the particle acceleration in most astrophysical accelerators. The most important factor is
the ratio of the particle scattering mean free path to the size of the source $\lambda/L\sim \tsc/\tcross$.
In what follows we will use the pitch angle averaged scattering time $\tsc\sim \lambda/v$ and
crossing time $\tcross\sim L/v$, where $v$ is the particle speed. In
sources with strong guiding magnetic field, the divergence or convergence of the field described by the parameter 
$\eta=B_{esc}/B_0$, where $B_{esc}$ denotes the increased field at the boundary, and $B_0$ is the
field strength in the center of the domain, also plays an important role. The third factor is the
momentum (or for
magnetized plasmas) the pitch angle distribution. In the strong diffusion limit this ratio is small
($\tsc\ll \tcross$) and the particles are isotropized quickly. They are able to random walk across
the
source with $\tesc\sim \tau_{\rm cross}^2/\tsc$ without much effect due to magnetic field variations
on the scale $h_b\equiv -B/(\partial B/\partial z\sim L\gg \lambda$.
On the other hand, in the weak diffusion limit with $\tsc\gg \tcross$, particles move freely
and escape within one crossing time unless there is a strong field convergence toward the boundary
of the region, which can trap particle by mirroring. In this case the
escape time is determined by how fast particles are scatted into the loss cone, in which
case for an isotropic distribution $\tesc \propto \tsc$, with the proportionality constant
increasing with increasing field convergence rate $\eta$.
The three regimes can be summarized as:
 \[ {\tesc\over \tcross}=  \left\{
 \begin{array}{lll}
      1\, & \mbox{if $\tsc\gg \tcross$, Free stream}\\ 
           \tcross/\tsc\, & \mbox{if $\tsc\ll \tcross$, Strong diffusion}\\
           \propto \tsc/\tcross\, & \mbox{if $\tsc\gg \tcross$, Converging field}
 \label{cases}
 \end{array}
      \right. \]
      
The combination of the first two as $\tesc/\tcross=1+\tcross/\tsc$ is commonly used \citep[see, e.g.][]{Petrosian-Liu-2004} for uniform or chaotic magnetic field situations. This can be
generalized by combining with the third case to a simple
analytical approximate formula relating the
particle escape and scattering as
\citep{Petrosian-2016}
\begin{equation}
\tesc= {\tau_{\text{cross}}}\left[C_1(\eta) +
C_2(\eta)\frac{\tau_{\text{cross}}}{\tau_{\text{sc}}} +
C_3(\eta)\frac{\tau_{\text{sc}}}{\tau_{\text{cross}}}\right] \,,
\label{eq:tauesc}
\end{equation}
with coefficients that  depend only on the value of $\eta$ and on the degree of the isotropy
of the distribution.
The appendix in \citet{Malyshkin-Kulsrud-2001} gives an extensive discussion on this
dependences leading to the following equation valid for an isotropic distribution.

\begin{equation}
\tesc =  {\tau_{\text{cross}}}\left[2\eta + \frac{\tau_{\text{cross}}}{\tau_{\text{sc}}}
+ \ln{\eta}\frac{\tau_{\text{sc}}}{\tau_{\text{cross}}}\right] \,.
\label{eq:ouresc}
\end{equation}
For distributions with substantial anisotropy we expect deviations from this equations,
especially in the weak diffusion limit. Below we compare our simulation results with this equation.

\section{Particle transport model with field-line convergence}
\label{model}

In this section we evaluate the effects of pitch angle scattering and field convergence on the
transport of particles through the acceleration site. For simplicity, we ignore any energy gain (often attributed to scattering by turbulence) or loss (as expected in Coulomb scatterings and radiative
processes) that are normally present in acceleration sites.

\subsection{Transport equation and coefficients}
\label{kineq}

To study the influence of pitch-angle scattering on the particle escape at a fixed given energy, the general Fokker-Planck equation for particle transport \citep[e.g.,][]{Schlickeiser-1989,Amstrong-etal-2012}, which is common in solar flare
\citep[e.g.,][]{Leach-Petrosian-1981,McTiernan-Petrosian-1991} and interplanetary particle transport studies \citep[e.g.,][]{Roelof-1969,Earl-1981,Effenberger-Litvinenko-2014}, can be reduced to the following
energy independent form:
\begin{equation}
  \frac{\partial f}{\partial t}
+ \mu v \frac{\partial f}{\partial z}
+ \frac{v}{2 h_B}(1-\mu^2)\frac{\partial f}{\partial \mu}
= \frac{\partial}{\partial \mu} \left( D_{\mu\mu}\frac{\partial f}{\partial \mu}\right), \,
\label{eq:PAFPE}
\end{equation}
for $f=f(z,\mu,t)$. Here, $z$ is the distance along the mean magnetic field $B$, $t$ is
time, $\mu$ is the cosine of the particle pitch angle, 
$h_B=-B/(\partial B/\partial z)$ is the 
field convergence scale height, and $D_{\mu\mu}$ denotes the
pitch-angle diffusion Fokker-Planck coefficient\footnote{Note that depending on the exact definition of $f$, the $h_B$ dependent convergence term may be written in implicit or explicit form \citep[e.g.][]{Earl-1981,Litvinenko-Noble-2013}. For our purposes, in the following, we will consider a simplified model of mirroring and confinement that doesn't require this term in the integration scheme, as described in the following sections.}.

We consider isotropic pitch-angle scattering with the diffusion coefficient
\begin{equation}
D_{\mu\mu}=D_0(1-\mu^2) \, ,
\label{eq:Dmumu}
\end{equation}
where $D_0$ is a constant which quantifies the strength of the scattering. In the diffusion approximation, which is essentially an
average of the pitch-angle dependence of the particles, the respective parallel spatial diffusion coefficient along $z$ is given by
\citep[e.g.,][]{Dung-Petrosian-1994,Schlickeiser-Shalchi-2008}
\begin{equation}
  \kappa_{zz} = \frac{v\lambda}{3} = \frac{v^2}{8}\int_{-1}^1 d\mu \frac{(1-\mu^2)^2}{D_{\mu\mu}} = \frac{v^2}{6D_0}
\label{eq:kappa}
\end{equation}
The scattering time can thus be defined as 
\begin{equation}
\tau_{\text{sc}} = \frac{\lambda}{v}= \frac{1}{2D_0} \,.
\label{eq:tausc}
\end{equation}

We normalize all quantities to the length of the system $L$ and the particle speed $v$. Time is thus measured in units of $\frac{L}{v}$, while we define the ensemble crossing time as $\tau_{\text{cross}}= \frac{L}{v\sqrt{2}}$, appropriate for an isotropic pitch-angle distribution.

\subsection{Stochastic simulation scheme}
\label{scheme}

\begin{figure*}[ht]
\noindent\includegraphics[width=0.99\textwidth]{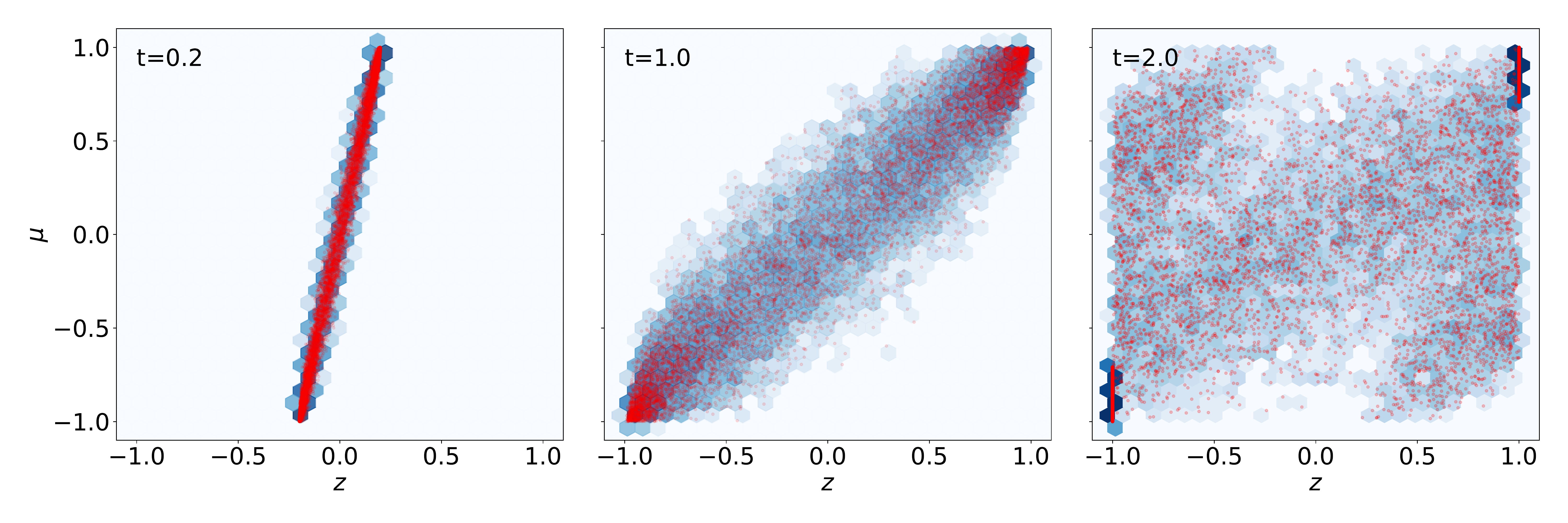}
\noindent\includegraphics[width=0.99\textwidth]{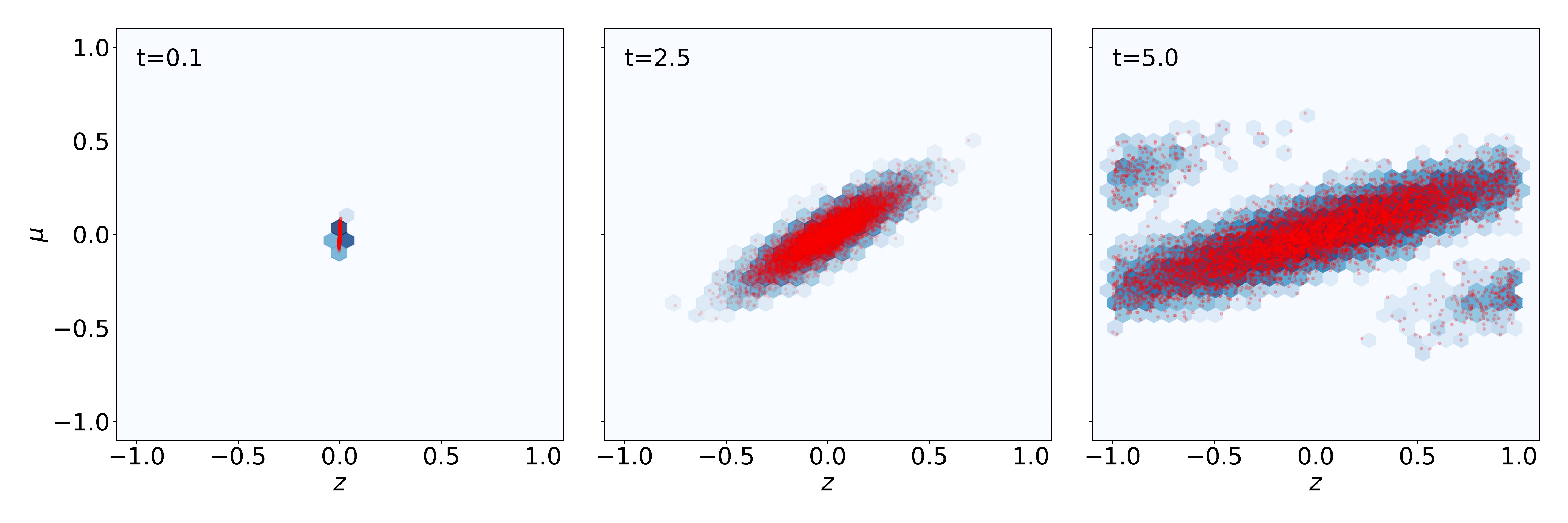}
\caption{Pseudo-particle distribution in $z-\mu$ phase space for isotropic (upper panel) and 'pancake' (lower panel) injection at three times during the evolution with $\eta=2$ and $D_0=0.1$, or  $\tau_{\rm sc}=5 L/v$, (upper panel) and
$D_0 = 0.003$ (lower panel). The red circles indicate individual particles and the blue background coloring is given by a hexagonal binning with arbitrary units (darker blue means more particles).}
\label{fig:iso-particles}
\end{figure*}

Stochastic differential equations (SDEs) are used in many contexts to solve Fokker-Planck type equations. In space physics, they are often employed to solve particle propagation problems, such as cosmic-ray
modulation \citep{Strauss-etal-2011,Effenberger-etal-2012}, SEP transport \citep{Droege-etal-2010}, shock acceleration \citep{Achterberg-Schure-2011,Zuo-etal-2011}, focused acceleration \citep{Amstrong-etal-2012},
and pick-up ion evolution \citep{Fichtner-etal-1996,Chalov-Fahr-1998}. For a recent account of numerical methods and other aspects connected to this
approach, see, e.g., \citet{Kopp-etal-2012} and the review \citet{Strauss-Effenberger-2017}. In the context of solar flares, \citet{MacKinnon-Craig-1991} presented one of the first simulation schemes based on the SDE approach. Recently, this method has also been employed in studies of the warm-target model \citep{Jeffrey-etal-2014,Kontar-etal-2015} and coupled hydrodynamic simulations of solar flares \citep[e.g.][]{Moravec-etal-2016}.

For our purposes, we can recast the transport equation (\ref{eq:PAFPE}) with isotropic scattering into the following set of SDEs \citep[e.g.,][]{Gardiner-2009}
\begin{align}
dz &= \mu v dt , \\
d\mu &= \left[\frac{v}{2 h_B}(1-\mu^2) - 2D_0\mu\right]dt + \sqrt{2D_0(1-\mu^2)}dW(t)\,,
\label{eq:sde}
\end{align}
where $W(t)$ represents a Wiener process with zero mean and variance $t$.

Ignoring the convergence term, which can be treated separately (see below), these equations can be solved numerically using a simple Euler approximation scheme \citep{Kloeden-Platen-1995}:
\begin{align}
z_{t+\Delta t} &= z_t + \mu_t v\Delta t , 
\label{eq:sdemil1} \\
\mu_{t+\Delta t} &=  \mu_t - 2D_0\mu_t\Delta t + \sqrt{2D_0(1-\mu_t^2)\Delta t}\epsilon_t \,,
\label{eq:sdemil2}
\end{align}
where $\epsilon_t$ is a normal random variable with zero mean and unit variance and $\Delta t$ is a small time step \citep{Strauss-Effenberger-2017}. We use reflecting boundaries at $\mu=\pm 1$ to conserve the probability.

In practice, this system of coupled ordinary SDEs is solved numerically by following a large
number of pseudo-particle orbits according to the above scheme and obtaining the distribution functions by corresponding averages over the particle positions in phase space. We can consider different initial conditions by changing the starting position of the particles using a suitable sampling of the initial distribution. For the purposes of this study, we will focus on three different forms of initial pitch-angle distributions: isotropic, 'pancake', i.e.\ sharply peaked at $90^\circ$ pitch-angle, and beam-like, i.e.\ two beams of particles at $\mu=\pm 1$.
\begin{figure*}[ht]
\noindent\includegraphics[width=0.99\textwidth]{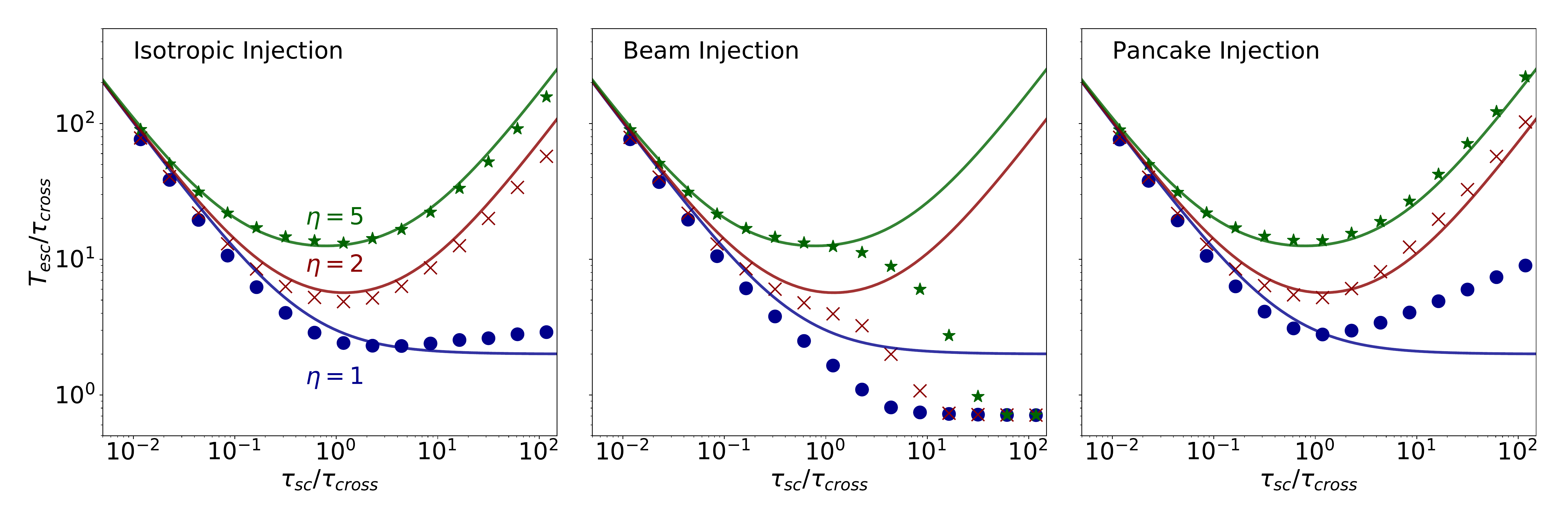}
\caption{Particle escape times vs. scattering time $\tau_{\text{sc}}$ for three different values of mirror strength: $\eta_1$=1 (blue), $\eta_2$=2 (red) and $\eta_3$=4 (green) and three different injection functions. Points are from our simulations and solid lines are based on the
analytic approximation of Equation~(\ref{eq:ouresc}).
}
\label{fig:escape_times_iso}
\end{figure*}

\subsection{Field convergence and particle escape}
\label{convergence}

The two main processes in our model are pitch-angle scattering and the background magnetic field
convergence or mirroring effect. The latter can be included in a straightforward way in the
SDE model described above. Consider a magnetic field increase prescribed by the parameter 
$\eta$ defined in \S \ref{escape}. Note that for a constant magnetic field, the scale height is given as $h_b=L\ln \eta$.
The loss cone of particles, i.e.\ the threshold pitch-angle, beyond which particles escape, can be defined as
$\mu_{\text{crit}}=\sqrt{1-1/\eta}$ \citep{Malyshkin-Kulsrud-2001}.

We assume an idealized field geometry, where the effect of field convergence only applies point-wise at the boundaries of
the domain. With this condition, particles are only
reflected back into the domain at the boundary ($z=\pm1$) if their pitch-angle is smaller than
$\mu_{\text{crit}}$. Otherwise they are counted as escaped particles. 
Thus, for a given set of parameters, the escape
time $T_{\text{esc}}$ can simply be calculated as the average time the particles take to
leave the domain through the loss cone. It is therefore not necessary, as mentioned before,
to include the convergence term in the integration scheme explicitly, since the effect is accounted
for by the selective reflection or escape of particles through the loss-cone. In other words, the
escape time is approximated as ensemble average or first moment of the residence time distribution
of escaping particles. In practice, the simulation runs until all particles have escape the domain.
In the upper panel of Figure~\ref{fig:iso-particles} it can be seen how particles exit the
phase-space domain at $z=\pm 1$ for $\mu$ close to $\pm 1$.

Furthermore, Figure~\ref{fig:iso-particles} illustrates the time evolution of the pseudo-particle ensemble and the associated phase-space distribution resulting from a hexagonal binning. In the upper panel, we consider an uniform initial pitch-angle distribution, $\eta=2$ and a diffusion
coefficient of $D_0 = 0.1$ (or $\tau_{\rm sc}=5 L/v)$. We see how particles that started close to $\mu=\pm1$ move quickly to the escape region and are not reflected back into the domain. At $t=2$ the contribution of reflected particles with $\mu<\mu_{\text{crit}}$ is visible as secondary patches in the distribution. Eventually, most of the particles will have escaped the region and this allows to calculate the approximate escape time. The lower panel shows the evolution for a 'pancake'-like injection with a sharp peak at $\mu=0$. We also reduce the diffusion coefficient to $D_0 = 0.003$ to illustrate the effect of weak diffusion. We find that particles are generally not able to reach the escape cone even after about 5 crossing times. Their effective crossing and residence time is significantly prolonged due to the combination of the initial condition being far away from the loss cone and the weak diffusion.

\section{Results and Discussion}
\label{results}

To study the expected behavior of the escape time based on the kinetic equations, and
to test the validity of the above equation, we have carried out simulations with $N=10000$
particles  with different values for $D_0$ and $\eta$ and for the three initial pitch-angle
distributions (isotropic, pancake, beam). Figure~\ref{fig:escape_times_iso} summarizes the results in
three panels for the three initial conditions. 

The left panel shows the case of isotropic injection, which as expected shows the best overall
agreement between the simulation results (symbols) and the
analytic expression
(Equation~\ref{eq:ouresc}) for all three different convergence parameters $\eta$. 
Note that in the strong diffusion regime, the numerical results converge to the random walk expression
$\tesc\sim \tcross^2/\tau_{\text{sc}}$ 
independent of $\eta$ and isotropy. However, there are significant deviations from this in the
intermediate regime (starting at lower values of $\tsc/\tcross$ for larger $\eta$). For weak 
diffusion, the $\eta=1$ case shows the free escape of
particles, while the other two cases exhibit the $\tesc\propto \tau_{\text{sc}}$ behavior
that is reproduced well by the simulations.

The middle panel shows the result for a beam injection of particles. With the forward SDE numerical scheme used here, we can inject particles at exactly $\mu=\pm 1$, resembling a very narrow bidirectional beam. The universal behavior in the
strong diffusion regime is reproduced again, while the the analytic expression breaks down in the weak 
diffusion regime, with all cases eventually converging to the crossing
time. This happens at larger scattering times for higher $\eta$ cases with smaller loss cones,
$\theta_{\rm lcone}\sim 1/\sqrt{\eta}$. This can easily be explained by the fact that at shorter
scattering times, particles initially in the loss cone are scattered out of it and remain in the
site for a longer time for higher values of $\eta$.

Finally, in the right panel, we find that an initially strongly peaked pitch-angle distribution near
$\mu=0$ (pancake) shows overall similar behavior, but as expected slightly
larger values for the escape time
than in the isotropic injection case. The most notable difference in the
weak diffusion regime for $\eta=1$ can be explained due to the difficulty of scattering
of particles from their initial large pitch angles ($\mu=0$) to small angles at longer and longer
scattering times.


\section{Summary \& Conclusions}
\label{summary}
In this study, we investigated the particle confinement and escape resulting from the interplay of
isotropic turbulent pitch-angle scattering and magnetic field convergence. We compared numerical
solutions for the relation between scattering and escape time calculated with an SDE scheme for the
particle transport equation with an analytical approximation formula, based on the work of \citet{Malyshkin-Kulsrud-2001}. We found good agreement
between the approximation and the simulation results but also notable differences in the weak
diffusion regime that depend both on the initial pitch-angle distribution of particles and the 
field convergence rate represented by the mirror ratio $\eta$ or scale height $h_B$.

The investigation of the acceleration and transport of particles in magnetized plasmas depends
crucially on the escape time, because in most situations we do not observe the particles at the
acceleration site. Instead, we observe the particles which have escaped the acceleration site and reached the near Earth instruments as
cosmic rays and SEPs or indirectly through the radiation they produce often away from the acceleration sites.

In some sources and under favorable observational situations, it is possible to measure the escape
time and its energy dependence \citep[see, e.g.][]{Chen-Petrosian-2013,Petrosian-Chen-2014}. The
relation we have established here can be used in such situations to determine the scattering time or the pitch-angle
diffusion coefficient and hence provide information on acceleration and transport mechanisms.
Expansion of these results to a more realistic situation that includes anisotropic
pitch-angle diffusion coefficients
and the dependence on energy of the processes discussed here can shed light on additional effects and the
other important diffusion coefficient, namely energy or momentum diffusion, that plays an equally
significant role in acceleration and transport processes. These aspects will be addressed in our
future works.

\begin{acknowledgements} 
This work was partially supported by NASA grants NNX14AG03G and NNX17AK25G. We acknowledge helpful discussions with J.\ McTiernan and N.\ Jeffrey and support from the International Space Science Institute through the ISSI team on "Solar flare acceleration signatures and their connection to solar energetic particles".
\end{acknowledgements} 

\newpage

\begin{thebibliography}{}
\expandafter\ifx\csname natexlab\endcsname\relax\def\natexlab#1{#1}\fi
\providecommand{\url}[1]{\href{#1}{#1}}
\providecommand{\dodoi}[1]{doi:~\href{http://doi.org/#1}{\nolinkurl{#1}}}
\providecommand{\doeprint}[1]{\href{http://ascl.net/#1}{\nolinkurl{http://ascl.net/#1}}}
\providecommand{\doarXiv}[1]{\href{https://arxiv.org/abs/#1}{\nolinkurl{https://arxiv.org/abs/#1}}}

\bibitem[{{Achterberg} \& {Schure}(2011)}]{Achterberg-Schure-2011}
{Achterberg}, A., \& {Schure}, K.~M. 2011, \mnras, 411, 2628,
  \dodoi{10.1111/j.1365-2966.2010.17868.x}

\bibitem[{{Ackermann} {et~al.}(2017){Ackermann}, {Allafort}, {Baldini},
  {Barbiellini}, {Bastieri}, {Bellazzini}, {Bissaldi}, {Bonino}, {Bottacini},
  {Bregeon}, {Bruel}, {Buehler}, {Cameron}, {Caragiulo}, {Caraveo},
  {Cavazzuti}, {Cecchi}, {Charles}, {Ciprini}, {Costanza}, {Cutini},
  {D'Ammando}, {de Palma}, {Desiante}, {Digel}, {Di Lalla}, {Di Mauro}, {Di
  Venere}, {Drell}, {Favuzzi}, {Fukazawa}, {Fusco}, {Gargano}, {Giglietto},
  {Giordano}, {Giroletti}, {Grenier}, {Guillemot}, {Guiriec}, {Jogler},
  {J{\'o}hannesson}, {Kashapova}, {Krucker}, {Kuss}, {La Mura}, {Larsson},
  {Latronico}, {Li}, {Liu}, {Longo}, {Loparco}, {Lubrano}, {Magill}, {Maldera},
  {Manfreda}, {Mazziotta}, {Mitthumsiri}, {Mizuno}, {Monzani}, {Morselli},
  {Moskalenko}, {Negro}, {Nuss}, {Ohsugi}, {Omodei}, {Orlando}, {Pal'shin},
  {Paneque}, {Perkins}, {Pesce-Rollins}, {Petrosian}, {Piron}, {Principe},
  {Rain{\`o}}, {Rando}, {Razzano}, {Reimer}, {Rubio da Costa}, {Sgr{\`o}},
  {Simone}, {Siskind}, {Spada}, {Spandre}, {Spinelli}, {Tajima}, {Thayer},
  {Torres}, {Troja}, \& {Vianello}}]{Ackermann-etal-2017}
{Ackermann}, M., {Allafort}, A., {Baldini}, L., {et~al.} 2017, \apj, 835, 219,
  \dodoi{10.3847/1538-4357/835/2/219}

\bibitem[{{Ajello} {et~al.}(2014){Ajello}, {Albert}, {Allafort}, {Baldini},
  {Barbiellini}, {Bastieri}, {Bellazzini}, {Bissaldi}, {Bonamente}, {Brandt},
  {Bregeon}, {Brigida}, {Bruel}, {Buehler}, {Buson}, {Caliandro}, {Cameron},
  {Caraveo}, {Cecchi}, {Charles}, {Chekhtman}, {Chiang}, {Chiaro}, {Ciprini},
  {Claus}, {Cohen-Tanugi}, {Cominsky}, {Conrad}, {Cutini}, {D'Ammando}, {de
  Palma}, {Dermer}, {Desiante}, {Digel}, {Silva}, {Drell}, {Drlica-Wagner},
  {Favuzzi}, {Focke}, {Franckowiak}, {Fukazawa}, {Fusco}, {Gargano},
  {Gasparrini}, {Germani}, {Giglietto}, {Giommi}, {Giordano}, {Giroletti},
  {Glanzman}, {Godfrey}, {Grenier}, {Grove}, {Guiriec}, {Hadasch}, {Hayashida},
  {Hays}, {Horan}, {Hou}, {Hughes}, {Inoue}, {Jackson}, {Jogler},
  {J{\'o}hannesson}, {Johnson}, {Johnson}, {Kamae}, {Kn{\"o}dlseder},
  {Kocevski}, {Kuss}, {Lande}, {Larsson}, {Latronico}, {Longo}, {Loparco},
  {Lott}, {Lovellette}, {Lubrano}, {Mayer}, {Mazziotta}, {McEnery},
  {Michelson}, {Mizuno}, {Moiseev}, {Monte}, {Monzani}, {Morselli},
  {Moskalenko}, {Murgia}, {Murphy}, {Nakamori}, {Nemmen}, {Nuss}, {Ohno},
  {Ohsugi}, {Omodei}, {Orienti}, {Orlando}, {Ormes}, {Paneque}, {Panetta},
  {Perkins}, {Pesce-Rollins}, {Petrosian}, {Piron}, {Pivato}, {Porter},
  {Rain{\`o}}, {Rando}, {Razzano}, {Reimer}, {Reimer}, {Roth}, {Schulz},
  {Sgr{\`o}}, {Siskind}, {Spandre}, {Spinelli}, {Takahashi}, {Thayer},
  {Thayer}, {Thompson}, {Tibaldo}, {Tinivella}, {Tosti}, {Troja}, {Usher},
  {Vandenbroucke}, {Vasileiou}, {Vianello}, {Vitale}, {Werner}, {Winer},
  {Wood}, {Wood}, \& {Yang}}]{Ajello-etal-2014}
{Ajello}, M., {Albert}, A., {Allafort}, A., {et~al.} 2014, \apj, 789, 20,
  \dodoi{10.1088/0004-637X/789/1/20}

\bibitem[{{Armstrong} {et~al.}(2012){Armstrong}, {Litvinenko}, \&
  {Craig}}]{Amstrong-etal-2012}
{Armstrong}, C.~K., {Litvinenko}, Y.~E., \& {Craig}, I.~J.~D. 2012, \apj, 757,
  165, \dodoi{10.1088/0004-637X/757/2/165}

\bibitem[{{Bian} {et~al.}(2014){Bian}, {Emslie}, {Stackhouse}, \&
  {Kontar}}]{Bian-etal-2014}
{Bian}, N.~H., {Emslie}, A.~G., {Stackhouse}, D.~J., \& {Kontar}, E.~P. 2014,
  \apj, 796, 142, \dodoi{10.1088/0004-637X/796/2/142}

\bibitem[{{Chalov} \& {Fahr}(1998)}]{Chalov-Fahr-1998}
{Chalov}, S.~V., \& {Fahr}, H.~J. 1998, \aap, 335, 746

\bibitem[{{Chen} \& {Petrosian}(2013)}]{Chen-Petrosian-2013}
{Chen}, Q., \& {Petrosian}, V. 2013, \apj, 777, 33,
  \dodoi{10.1088/0004-637X/777/1/33}

\bibitem[{{Dr{\"o}ge} {et~al.}(2010){Dr{\"o}ge}, {Kartavykh}, {Klecker}, \&
  {Kovaltsov}}]{Droege-etal-2010}
{Dr{\"o}ge}, W., {Kartavykh}, Y.~Y., {Klecker}, B., \& {Kovaltsov}, G.~A. 2010,
  \apj, 709, 912, \dodoi{10.1088/0004-637X/709/2/912}

\bibitem[{{Dung} \& {Petrosian}(1994)}]{Dung-Petrosian-1994}
{Dung}, R., \& {Petrosian}, V. 1994, \apj, 421, 550, \dodoi{10.1086/173670}

\bibitem[{{Earl}(1981)}]{Earl-1981}
{Earl}, J.~A. 1981, \apj, 251, 739, \dodoi{10.1086/159518}

\bibitem[{{Effenberger} {et~al.}(2012){Effenberger}, {Fichtner}, {Scherer},
  {Barra}, {Kleimann}, \& {Strauss}}]{Effenberger-etal-2012}
{Effenberger}, F., {Fichtner}, H., {Scherer}, K., {et~al.} 2012, \apj, 750,
  108, \dodoi{10.1088/0004-637X/750/2/108}

\bibitem[{{Effenberger} \& {Litvinenko}(2014)}]{Effenberger-Litvinenko-2014}
{Effenberger}, F., \& {Litvinenko}, Y.~E. 2014, \apj, 783, 15,
  \dodoi{10.1088/0004-637X/783/1/15}

\bibitem[{{Effenberger} {et~al.}(2017){Effenberger}, {Rubio da Costa}, {Oka},
  {Saint-Hilaire}, {Liu}, {Petrosian}, {Glesener}, \&
  {Krucker}}]{Effenberger-etal-2017}
{Effenberger}, F., {Rubio da Costa}, F., {Oka}, M., {et~al.} 2017, \apj, 835,
  124, \dodoi{10.3847/1538-4357/835/2/124}

\bibitem[{{Effenberger} {et~al.}(2016){Effenberger}, {Rubio da Costa}, \&
  {Petrosian}}]{Effenberger-etal-2016}
{Effenberger}, F., {Rubio da Costa}, F., \& {Petrosian}, V. 2016, in Journal of
  Physics Conference Series, Vol. 767, Journal of Physics Conference Series,
  012005

\bibitem[{{Fichtner} {et~al.}(1996){Fichtner}, {Le Roux}, {Mall}, \&
  {Rucinski}}]{Fichtner-etal-1996}
{Fichtner}, H., {Le Roux}, J.~A., {Mall}, U., \& {Rucinski}, D. 1996, \aap,
  314, 650

\bibitem[{{Fletcher}(1995)}]{Fletcher-1995}
{Fletcher}, L. 1995, \aap, 303, L9

\bibitem[{{Fletcher} \& {Martens}(1998)}]{Fletcher-Martens-1998}
{Fletcher}, L., \& {Martens}, P.~C.~H. 1998, \apj, 505, 418,
  \dodoi{10.1086/306137}

\bibitem[{{Gardiner}(2009)}]{Gardiner-2009}
{Gardiner}, C.~W. 2009, {Stochastic Methods: A Handbook for the Natural and
  Social Sciences} (Berlin: Springer)

\bibitem[{{Jeffrey} {et~al.}(2014){Jeffrey}, {Kontar}, {Bian}, \&
  {Emslie}}]{Jeffrey-etal-2014}
{Jeffrey}, N.~L.~S., {Kontar}, E.~P., {Bian}, N.~H., \& {Emslie}, A.~G. 2014,
  \apj, 787, 86, \dodoi{10.1088/0004-637X/787/1/86}

\bibitem[{{Jin} {et~al.}(2018){Jin}, {Petrosian}, {Liu}, {Nitta}, {Omodei},
  {Rubio da Costa}, {Effenberger}, {Li}, {Pesce-Rollins}, {Allafort}, \&
  {Manchester}}]{Jin-etal-2018}
{Jin}, M., {Petrosian}, V., {Liu}, W., {et~al.} 2018, ApJ, in press.
\newblock \doarXiv{1807.01427}

\bibitem[{{Ka{\v s}parov{\'a}} \&
  {Karlick{\'y}}(2009)}]{Kasparova-Karlicky-2009}
{Ka{\v s}parov{\'a}}, J., \& {Karlick{\'y}}, M. 2009, \aap, 497, L13,
  \dodoi{10.1051/0004-6361/200911898}

\bibitem[{{Kloeden} \& {Platen}(1995)}]{Kloeden-Platen-1995}
{Kloeden}, P., \& {Platen}, E. 1995, {Numerical methods for stochastic
  differential equations} (Spinger, Berlin)

\bibitem[{{Kontar} {et~al.}(2014){Kontar}, {Bian}, {Emslie}, \&
  {Vilmer}}]{Kontar-etal-2014}
{Kontar}, E.~P., {Bian}, N.~H., {Emslie}, A.~G., \& {Vilmer}, N. 2014, \apj,
  780, 176, \dodoi{10.1088/0004-637X/780/2/176}

\bibitem[{{Kontar} {et~al.}(2015){Kontar}, {Jeffrey}, {Emslie}, \&
  {Bian}}]{Kontar-etal-2015}
{Kontar}, E.~P., {Jeffrey}, N.~L.~S., {Emslie}, A.~G., \& {Bian}, N.~H. 2015,
  \apj, 809, 35, \dodoi{10.1088/0004-637X/809/1/35}

\bibitem[{{Kopp} {et~al.}(2012){Kopp}, {B{\"u}sching}, {Strauss}, \&
  {Potgieter}}]{Kopp-etal-2012}
{Kopp}, A., {B{\"u}sching}, I., {Strauss}, R.~D., \& {Potgieter}, M.~S. 2012,
  Computer Physics Communications, 183, 530, \dodoi{10.1016/j.cpc.2011.11.014}

\bibitem[{{Krucker} {et~al.}(2010){Krucker}, {Hudson}, {Glesener}, {White},
  {Masuda}, {Wuelser}, \& {Lin}}]{Krucker-etal-2010}
{Krucker}, S., {Hudson}, H.~S., {Glesener}, L., {et~al.} 2010, \apj, 714, 1108,
  \dodoi{10.1088/0004-637X/714/2/1108}

\bibitem[{{Krucker} \& {Lin}(2008)}]{Krucker-Lin-2008}
{Krucker}, S., \& {Lin}, R.~P. 2008, \apj, 673, 1181, \dodoi{10.1086/524010}

\bibitem[{{Krucker} {et~al.}(2007){Krucker}, {White}, \&
  {Lin}}]{Krucker-etal-2007}
{Krucker}, S., {White}, S.~M., \& {Lin}, R.~P. 2007, \apjl, 669, L49,
  \dodoi{10.1086/523759}

\bibitem[{{Leach} \& {Petrosian}(1981)}]{Leach-Petrosian-1981}
{Leach}, J., \& {Petrosian}, V. 1981, \apj, 251, 781, \dodoi{10.1086/159521}

\bibitem[{{Leach} \& {Petrosian}(1983)}]{Leach-Petrosian-1983}
---. 1983, \apj, 269, 715, \dodoi{10.1086/161081}

\bibitem[{{Lin} {et~al.}(2002){Lin}, {Dennis}, {Hurford}, {Smith}, {Zehnder},
  {Harvey}, {Curtis}, {Pankow}, {Turin}, {Bester}, {Csillaghy}, {Lewis},
  {Madden}, {van Beek}, {Appleby}, {Raudorf}, {McTiernan}, {Ramaty}, {Schmahl},
  {Schwartz}, {Krucker}, {Abiad}, {Quinn}, {Berg}, {Hashii}, {Sterling},
  {Jackson}, {Pratt}, {Campbell}, {Malone}, {Landis}, {Barrington-Leigh},
  {Slassi-Sennou}, {Cork}, {Clark}, {Amato}, {Orwig}, {Boyle}, {Banks},
  {Shirey}, {Tolbert}, {Zarro}, {Snow}, {Thomsen}, {Henneck}, {McHedlishvili},
  {Ming}, {Fivian}, {Jordan}, {Wanner}, {Crubb}, {Preble}, {Matranga}, {Benz},
  {Hudson}, {Canfield}, {Holman}, {Crannell}, {Kosugi}, {Emslie}, {Vilmer},
  {Brown}, {Johns-Krull}, {Aschwanden}, {Metcalf}, \& {Conway}}]{Lin-etal-2002}
{Lin}, R.~P., {Dennis}, B.~R., {Hurford}, G.~J., {et~al.} 2002, \solphys, 210,
  3, \dodoi{10.1023/A:1022428818870}

\bibitem[{{Litvinenko} \& {Noble}(2013)}]{Litvinenko-Noble-2013}
{Litvinenko}, Y.~E., \& {Noble}, P.~L. 2013, \apj, 765, 31,
  \dodoi{10.1088/0004-637X/765/1/31}

\bibitem[{{Liu} {et~al.}(2013){Liu}, {Chen}, \& {Petrosian}}]{Liu-etal-2013}
{Liu}, W., {Chen}, Q., \& {Petrosian}, V. 2013, \apj, 767, 168,
  \dodoi{10.1088/0004-637X/767/2/168}

\bibitem[{{Liu} {et~al.}(2006){Liu}, {Liu}, {Jiang}, \&
  {Petrosian}}]{Liu-etal-2006}
{Liu}, W., {Liu}, S., {Jiang}, Y.~W., \& {Petrosian}, V. 2006, \apj, 649, 1124,
  \dodoi{10.1086/506268}

\bibitem[{{Liu} {et~al.}(2008){Liu}, {Petrosian}, {Dennis}, \&
  {Jiang}}]{Liu-etal-2008}
{Liu}, W., {Petrosian}, V., {Dennis}, B.~R., \& {Jiang}, Y.~W. 2008, \apj, 676,
  704, \dodoi{10.1086/527538}

\bibitem[{{MacKinnon} \& {Craig}(1991)}]{MacKinnon-Craig-1991}
{MacKinnon}, A.~L., \& {Craig}, I.~J.~D. 1991, \aap, 251, 693

\bibitem[{{Malyshkin} \& {Kulsrud}(2001)}]{Malyshkin-Kulsrud-2001}
{Malyshkin}, L., \& {Kulsrud}, R. 2001, \apj, 549, 402, \dodoi{10.1086/319080}

\bibitem[{{Masuda} {et~al.}(1994){Masuda}, {Kosugi}, {Hara}, {Tsuneta}, \&
  {Ogawara}}]{Masuda-etal-1994}
{Masuda}, S., {Kosugi}, T., {Hara}, H., {Tsuneta}, S., \& {Ogawara}, Y. 1994,
  \nat, 371, 495, \dodoi{10.1038/371495a0}

\bibitem[{{McTiernan} \& {Petrosian}(1991)}]{McTiernan-Petrosian-1991}
{McTiernan}, J.~M., \& {Petrosian}, V. 1991, \apj, 379, 381,
  \dodoi{10.1086/170513}

\bibitem[{{Moravec} {et~al.}(2016){Moravec}, {Varady}, {Ka{\v s}parov{\'a}}, \&
  {Kramoli{\v s}}}]{Moravec-etal-2016}
{Moravec}, Z., {Varady}, M., {Ka{\v s}parov{\'a}}, J., \& {Kramoli{\v s}}, D.
  2016, Astronomische Nachrichten, 337, 1020, \dodoi{10.1002/asna.201612427}

\bibitem[{{Oka} {et~al.}(2013){Oka}, {Ishikawa}, {Saint-Hilaire}, {Krucker}, \&
  {Lin}}]{Oka-etal-2013}
{Oka}, M., {Ishikawa}, S., {Saint-Hilaire}, P., {Krucker}, S., \& {Lin}, R.~P.
  2013, \apj, 764, 6, \dodoi{10.1088/0004-637X/764/1/6}

\bibitem[{{Oka} {et~al.}(2015){Oka}, {Krucker}, {Hudson}, \&
  {Saint-Hilaire}}]{Oka-etal-2015}
{Oka}, M., {Krucker}, S., {Hudson}, H.~S., \& {Saint-Hilaire}, P. 2015, \apj,
  799, 129, \dodoi{10.1088/0004-637X/799/2/129}

\bibitem[{{Oka} {et~al.}(2018){Oka}, {Birn}, {Battaglia}, {Chaston}, {Hatch},
  {Livadiotis}, {Imada}, {Miyoshi}, {Kuhar}, {Effenberger}, {Eriksson},
  {Khotyaintsev}, \& {Retin{\`o}}}]{Oka-etal-2018}
{Oka}, M., {Birn}, J., {Battaglia}, M., {et~al.} 2018, ArXiv e-prints.
\newblock \doarXiv{1805.09278}

\bibitem[{{Pesce-Rollins} {et~al.}(2015){Pesce-Rollins}, {Omodei}, {Petrosian},
  {Liu}, {Rubio da Costa}, {Allafort}, \& {Chen}}]{Pesce-Rollins-etal-2015}
{Pesce-Rollins}, M., {Omodei}, N., {Petrosian}, V., {et~al.} 2015, \apjl, 805,
  L15, \dodoi{10.1088/2041-8205/805/2/L15}

\bibitem[{{Petrosian}(2012)}]{Petrosian-2012}
{Petrosian}, V. 2012, \ssr, 173, 535, \dodoi{10.1007/s11214-012-9900-6}

\bibitem[{{Petrosian}(2016)}]{Petrosian-2016}
---. 2016, \apj, 830, 28, \dodoi{10.3847/0004-637X/830/1/28}

\bibitem[{{Petrosian} \& {Chen}(2010)}]{Petrosian-Chen-2010}
{Petrosian}, V., \& {Chen}, Q. 2010, \apjl, 712, L131,
  \dodoi{10.1088/2041-8205/712/2/L131}

\bibitem[{{Petrosian} \& {Chen}(2014)}]{Petrosian-Chen-2014}
---. 2014, \prd, 89, 103007, \dodoi{10.1103/PhysRevD.89.103007}

\bibitem[{{Petrosian} \& {Donaghy}(1999)}]{Petrosian-Donaghy-1999}
{Petrosian}, V., \& {Donaghy}, T.~Q. 1999, \apj, 527, 945,
  \dodoi{10.1086/308133}

\bibitem[{{Petrosian} {et~al.}(2002){Petrosian}, {Donaghy}, \&
  {McTiernan}}]{Petrosian-etal-2002}
{Petrosian}, V., {Donaghy}, T.~Q., \& {McTiernan}, J.~M. 2002, \apj, 569, 459,
  \dodoi{10.1086/339240}

\bibitem[{{Petrosian} \& {East}(2008)}]{Petrosian-East-2008}
{Petrosian}, V., \& {East}, W.~E. 2008, \apj, 682, 175, \dodoi{10.1086/588424}

\bibitem[{{Petrosian} \& {Kang}(2015)}]{Petrosian-Kang-2015}
{Petrosian}, V., \& {Kang}, B. 2015, \apj, 813, 5,
  \dodoi{10.1088/0004-637X/813/1/5}

\bibitem[{{Petrosian} \& {Liu}(2004)}]{Petrosian-Liu-2004}
{Petrosian}, V., \& {Liu}, S. 2004, \apj, 610, 550, \dodoi{10.1086/421486}

\bibitem[{{Piana} {et~al.}(2007){Piana}, {Massone}, {Hurford}, {Prato},
  {Emslie}, {Kontar}, \& {Schwartz}}]{Piana-etal-2007}
{Piana}, M., {Massone}, A.~M., {Hurford}, G.~J., {et~al.} 2007, \apj, 665, 846,
  \dodoi{10.1086/519518}

\bibitem[{{Roelof}(1969)}]{Roelof-1969}
{Roelof}, E.~C. 1969, in Lectures in High-Energy Astrophysics, ed.
  H.~{{\"O}gelman} \& J.~R. {Wayland}, 111

\bibitem[{{Schlickeiser}(1989)}]{Schlickeiser-1989}
{Schlickeiser}, R. 1989, \apj, 336, 243, \dodoi{10.1086/167009}

\bibitem[{{Schlickeiser} \& {Shalchi}(2008)}]{Schlickeiser-Shalchi-2008}
{Schlickeiser}, R., \& {Shalchi}, A. 2008, \apj, 686, 292,
  \dodoi{10.1086/591237}

\bibitem[{{Strauss} {et~al.}(2011){Strauss}, {Potgieter}, {B{\"u}sching}, \&
  {Kopp}}]{Strauss-etal-2011}
{Strauss}, R.~D., {Potgieter}, M.~S., {B{\"u}sching}, I., \& {Kopp}, A. 2011,
  \apj, 735, 83, \dodoi{10.1088/0004-637X/735/2/83}

\bibitem[{{Strauss} \& {Effenberger}(2017)}]{Strauss-Effenberger-2017}
{Strauss}, R.~D.~T., \& {Effenberger}, F. 2017, \ssr,
  \dodoi{10.1007/s11214-017-0351-y}

\bibitem[{{Zuo} {et~al.}(2011){Zuo}, {Zhang}, {Gamayunov}, {Rassoul}, \&
  {Luo}}]{Zuo-etal-2011}
{Zuo}, P., {Zhang}, M., {Gamayunov}, K., {Rassoul}, H., \& {Luo}, X. 2011,
  \apj, 738, 168, \dodoi{10.1088/0004-637X/738/2/168}

\end{thebibliography}

\end{document}